\documentclass[conference]{IEEEtran}
\IEEEoverridecommandlockouts
\usepackage{cite}
\usepackage{url}
\usepackage[T1]{fontenc}
\usepackage{amsmath,amssymb,amsfonts}
\usepackage{algorithmic}
\usepackage{graphicx}
\usepackage{textcomp}
\usepackage{xcolor}
\usepackage{booktabs}
\usepackage[ruled,vlined]{algorithm2e}
\usepackage{tikz}

\newcommand{\inlinebox}[1]{%
  \tikz[baseline=(X.base)]
    \node[
      draw=blue!50,
      fill=blue!10,
      rectangle,
      rounded corners=2pt,
      inner sep=2pt
    ] (X) {#1};%
}

\def\BibTeX{{\rm B\kern-.05em{\sc i\kern-.025em b}\kern-.08em
    T\kern-.1667em\lower.7ex\hbox{E}\kern-.125emX}}
\begin{document}

\title{Digital Twin-Based Simulation for Predictive Decision-Making in Waterway Logistics\\
\thanks{\IEEEauthorrefmark{1}Both authors contributed equally to this work.}
}

\author{\IEEEauthorblockN{Matthijs Jansen op de Haar\IEEEauthorrefmark{1}}
\IEEEauthorblockA{\textit{ETH Zürich}\\
Zürich, Switzerland \\
0009-0005-1673-592X}
\and
\IEEEauthorblockN{Daniel Frutos Rodriguez\IEEEauthorrefmark{1}}
\IEEEauthorblockA{\textit{Nanyang Technological University}\\
Singapore \\
0009-0002-0272-1568}
}

\maketitle

\begin{abstract}
This paper investigates the potential of a \emph{Digital Twin} (DT) for freight routing across inland waterway networks under uncertain water-level conditions. Existing approaches insufficiently account for increasing climate-induced volatility in water levels, which often result in higher operational costs and emissions due to the need for more expensive transport alternatives, such as road transport. These existing methods often rely on reactive countermeasures to remain resilient.

To address these limitations, six interviews with experts in domains related to inland shipping were conducted to identify three common contingency scenarios and appropriate operational responses. These scenarios were subsequently incorporated into a time-sliced simulation environment in which predictive decision-making, enabled by a DT environment, was compared against reactive approaches. The results demonstrate that predictive modeling substantially reduces operational costs and modal shifts at prediction accuracies between 70\% and 100\%, despite extreme conditions. In addition, the predictive model achieves an average 28.3\% reduction in fuel-related costs by reducing the total distance ships travel. The simulation outcomes were evaluated together with domain experts to assess the practical relevance and applicability of the proposed DT-enabled approach.
\end{abstract}


\section{Introduction}
In recent years, periods of low water levels in inland shipping have become increasingly frequent due to the progression of climate change \cite{Cascading_effects_sustained_low_water}. These low water levels significantly impact the load capacity of inland vessels \cite{effect_low_water_on_load_capacity}, often forcing operators to reduce cargo per trip, or completely inhibiting traffic altogether. As a consequence, more expensive transport alternatives, such as road transport, must be utilized. Previous studies have shown that such modal shifts lead to higher operational costs and increased emissions \cite{inland_transport_vs_road}.

To mitigate these challenges, many logistics organizations rely on a combination of publicly available and privately collected data to monitor and predict inland water levels. Given that inland waterways are used by a diverse set of stakeholders, including businesses, governmental agencies, and private stakeholders, close collaboration is essential for their effective operation. One example of such a collaboration is the \emph{Port of Twente (PoT)}\footnote{\url{https://portoftwente.com/}}, which serves as the basis for our study. The PoT forms a consortium of large corporations, research institutes, and governmental bodies in the Twente region of the Netherlands.

The PoT faces a significant challenge in integrating the heterogeneous data generated by its various stakeholders, many of which directly influence inland water levels through infrastructure such as locks. Successfully combining these fragmented data sources is essential for improving transparency into freight flows and water conditions across the inland waterway network. This naturally motivates the introduction of a \emph{Digital Twin} (DT), specifically in the form of a digital shadow, capable of integrating real-time and predictive information to support adaptive routing and operational decision-making.

Existing research has demonstrated substantial progress in water-level prediction and monitoring systems. However, the integration of these capabilities into DT systems that actively support operational decision-making remains limited. Current approaches largely focus on monitoring and visualization, while empirical validation of DT-supported decision-making in inland freight logistics remains scarce. In particular, little research examines how predictive approaches translate into operational responses such as rerouting, modal shifts, or cargo redistribution under uncertainty, or how predictive accuracy affects decision-making effectiveness. This study addresses these gaps by evaluating how predictive decision-making within a DT-supported simulation environment can improve resilience in inland freight logistics.

\section{Background}

\textbf{Climate-Induced Challenges in Waterway Logistics.} Climate change has led to increasingly frequent and severe low-water periods across European inland waterways, particularly along the Rhine and surrounding Dutch river systems that support the broader Benelux logistics network \cite{schweighofer2013}. These disruptions reduce vessel loading capacity, increase transport costs, and negatively affect freight throughput and supply chain reliability. Previous studies estimate that prolonged low-water disruptions can significantly reduce monthly throughput, while future economic losses may increase substantially if resilience is not improved \cite{vinke2022}. Additionally, periods of low water have already led to limited modal shifts from inland shipping to road and rail transport, thereby increasing costs and emissions \cite{jonkeren2013}. The growing use of larger vessels has further increased the vulnerability of inland shipping systems to drought-related disruptions, particularly in regions with extensive inland waterway infrastructure \cite{nemethy2022}.

\textbf{Operational and Infrastructure Vulnerabilities.} Recent studies show that the vulnerability of inland waterway transport to critical water levels has increased considerably in recent years \cite{bedoya2024,vinke2019}. Besides low water itself, critical infrastructure such as locks, bridges, and terminals also represent major operational bottlenecks during disruptions. Several European ports have been experiencing partial shutdowns or operational limitations due to unpredictable water levels and infrastructure constraints \cite{izaguirre2020}. As a result, users of the inland waterways increasingly require mechanisms for anticipating both gradual and sudden disruptions across the logistics network.

\textbf{Predictive Approaches and Decision-Support Systems.} To improve water-level predictability, both machine learning and traditional statistical forecasting approaches have been explored extensively. Artificial neural networks (ANN), long short-term memory (LSTM) models, ARIMA models, and hybrid forecasting approaches have all shown promising results for predicting inland water levels and hydrological behavior \cite{agaj2024,meissner2017}. In practice, operational systems combine real-time vessel sensor data with hydrological forecasting to improve situational awareness \cite{vandermark2020}. At the same time, simulation and optimization approaches such as discrete event simulation and heuristic routing models have been investigated to study disruptions and evaluate contingency responses \cite{delgado2020,shobayo2024}. Many existing industry tools integrate IoT-based monitoring, dashboards, and alert systems to visualize current and predicted water conditions \cite{pan2020}. However, these approaches often focus on prediction or visualization independently, rather than combining predictive insights with simulation-driven decision support and proactive response evaluation.

\textbf{Digital Twins in Logistics and Waterway Management.} DTs have been applied across a wide range of domains, including manufacturing, engineering, medicine, and smart agriculture \cite{verdouw2021,masison2021}. Existing DT frameworks commonly emphasize characteristics such as scalability, life cycle management, interoperability, and socio-technical integration \cite{autiosalo2020}. Within transport and logistics, DTs are increasingly used for real-time monitoring, simulation, situational awareness, and operational forecasting through integration of IoT and predictive models \cite{klar2023}. Prior research suggests that DTs can improve vessel scheduling, fleet management, coordination, and infrastructure planning while reducing delays, costs, and emissions \cite{eom2023,moshood2021}. Nevertheless, current DT implementations remain focused on monitoring and visualization rather than actively supporting strategic and operational decision-making under uncertainty.

\section{Methodology}
In this section, we will first identify relevant disruptive scenarios and their corresponding response strategies, based on expert discussions. Moreover, to assess the value of predictive decision-making enabled by DTs, we propose a simulation model of an inland waterway network. This simulation model quantifies how predictive accuracy affects operational outcomes. These outcomes are then compared against a reactive approach, in which no DT is available to support operational decision-making. Finally, we evaluate the suggested DT approach using an extensive expert study.

\begin{figure}[h]
    \centering
    \includegraphics[width=\columnwidth]{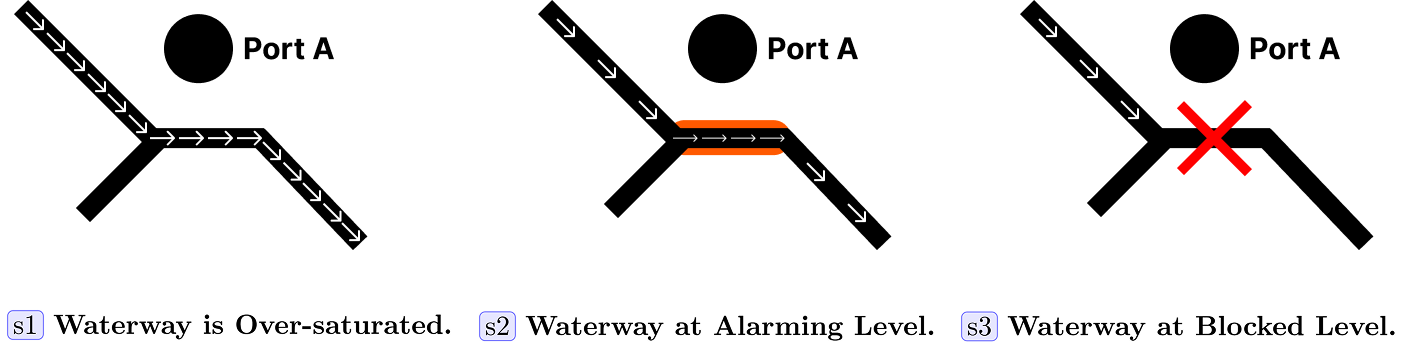}
    \caption{Inland waterway scenarios}
    \label{fig:scenarios}
\end{figure}

\textbf{Scenarios and Best Responses.}
Freight logistics can experience a range of undesirable outcomes and operational disruptions. Through extensive discussions with two experts in inland shipping, we identified three primary scenarios together with their corresponding optimal response strategies. These scenarios are presented in Figure \ref{fig:scenarios}.

\inlinebox{s1} \textit{Waterway is Over-saturated.} A waterway becomes oversaturated when the number of vessels operating within a particular section exceeds its available capacity. In such situations, traffic density reaches a level at which efficient and safe navigation can no longer be guaranteed. The most effective response is to reroute additional vessels through alternative waterways to reduce congestion and restore operational flow.

\inlinebox{s2} \textit{Waterway at Alarming Water Levels.} A waterway reaches an alarming level when the water depth within a particular section falls below or above a critical threshold. Under these conditions, larger vessels can no longer safely navigate the affected section. The most appropriate response is therefore to transfer freight to smaller vessels with reduced depth. However, this redistribution of traffic may subsequently contribute to waterway over-saturation, as described in \inlinebox{s1}, potentially requiring additional rerouting measures.

\inlinebox{s3} \textit{Waterway at Blocked Water Levels.} When water levels shift further beyond the initial critical threshold, they may eventually reach a point at which navigation is no longer possible for any vessel. Such conditions can indirectly contribute to waterway over-saturation, as described in \inlinebox{s1}, due to redistribution of traffic across alternative routes. In more severe cases, an entire connection between two arbitrary ports may become inaccessible. Under these circumstances, the most appropriate response is to shift freight transport to alternative modalities, such as road transport. This presents the most undesirable scenario as costs are drastically increased.

\textbf{Simulation Setup.} The simulation model captures the evolution of water levels across the inland waterway network over time. The underlying model is a mathematical abstraction, constructed through expert discussions. We subsequently evaluate this model in the context of a DT, using the notion of a digital shadow, assuming that baseline operations are reactive rather than predictive. In other words, without a DT, decisions are made after changes in water levels are observed. All model parameters are summarized in Table 1.

The network is represented as a graph \(G = (N, E)\), where nodes \(n \in N\) correspond to ports and locks, and edges \(e \in E\) correspond to waterway segments connecting these nodes. The model evolves over discrete time steps \(t = 0,1,\dots, T\). A fixed number of vessels \(s \in \mathbb{N}\) operate on the network, each assigned a randomly selected origin and destination pair consisting of two distinct nodes. For each edge \(e\), the water level at time \(t\) is denoted by \(h_e(t)\). The evolution of water levels follows a bounded random walk of the form: 
\begin{equation}
    h_e(t+1) = h_e(t) + \epsilon_e(t)
\end{equation}

Where \(\epsilon_e(t)\) is a zero-mean random variable with \(|\epsilon_e(t)| \leq S\). The parameter \(S\) controls the maximum fluctuation per timestep and serves as an entropy parameter governing variability in water levels. The model also includes a sine function that determines the mean water level and introduces seasonality, resulting in a decrease in water levels over time. This is important as we evaluate the resilience of the different models in the event of a contingency. While the evaluated scenarios focus on decreasing water levels, the underlying threshold-based structure of the model is symmetric and could similarly represent situations where water levels exceed operational thresholds, such as flooding or excessively high water conditions. Both the predictive and reactive models optimize for a reduction in modal shifts, as these incur the highest cost.

\begin{table}[h]
\caption{Model Parameters}
\label{tab:lnbip_example}
\centering

\setlength{\tabcolsep}{8pt}      
\renewcommand{\arraystretch}{1.15} 

\begin{tabular}{cccc}
\toprule
\textbf{Parameter} & \textbf{Description} & \textbf{Parameter} & \textbf{Description} \\
\midrule
\(t\) & \(time\ steps\) & \(a\) & \(accuracy\) \\
\(n\) & \(\#\ nodes\) & \(S\) & \(entropy\)  \\
\(e\) & \(\#\ edges\) & \(h\) & \(water\ level\) \\
\(s\) & \(\#\ ships\) & \(u\) & \(threshold\) \\
\(c\) & \(capacity\) & \(r\) & \(random\ seed\) \\
\end{tabular}
\end{table}

Each edge \(e\) has a finite capacity \(c_e \in \mathbb{N}\), such that the number of vessels assigned to that edge at time \(t\), denoted by \(f_e(t)\), satisfies \(f_e(t) \leq c_e\). When water levels fall below a predefined threshold \(u\), the scenarios in Figure \ref{fig:scenarios} can occur, resulting in rule-based optimal responses. The DT generates predictions of water levels with accuracy \(a \in [0,1]\) at timestep \(t+1\), enabling proactive measures such as rerouting or modal shifts. In the absence of a DT, responses occur only after threshold violations are observed, either resulting in a modal shift or in rerouting given sufficient capacity, assuming that these responses can be realized at any node.

\section{Simulation}
\begin{figure*}[t]
    \centering
    \includegraphics[width=\textwidth]{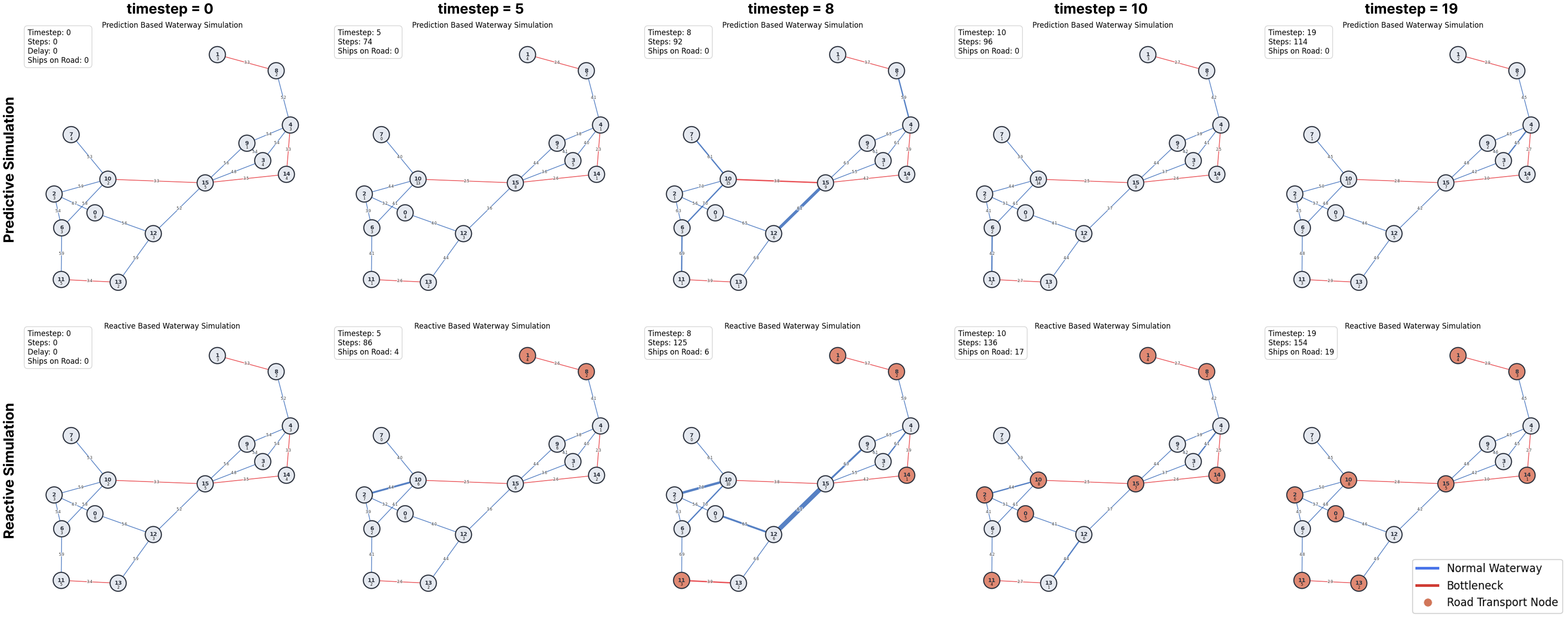}
    \caption{Simulation time steps over a predictive and reactive approach}
    \label{fig:simulation}
\end{figure*}

To evaluate the simulation, we investigate a randomly generated hub-and-spoke graph with 50 ships that resembles real-world inland waterways. Figure \ref{fig:simulation} shows one such instance of a model that has predictive accuracy of 90\% and a reactive model that can only make decisions based on present observations. In this particular simulation, approximately 25\% of the waterways become bottlenecks over time (i.e., the red edges), which is determined through an arbitrary threshold \(u =3\). 

We see that over time, the reactive model uses modal shifts in the red nodes. As the simulation progresses, the reactive model is unable to account for the reduction in water levels, particularly in isolated nodes. In contrast, as bottlenecks occurred, the predictive model was able to solve these by re-routing ships or shifting to smaller vessels in advance. For this reason, the simulation terminates at an earlier stage for the reactive approach, as there are no delays through re-routing or changing vessels. The predictive model terminates at \(timestep=19\), while the reactive model terminates at \(timestep=14\). Interestingly, due to route optimization, this does result in ships covering less distance and therefore using less fuel, as the reactive model has to use ad-hoc re-routing, resulting in sub-optimal routes. This means that beyond a reduction in modal shifts, there is also a cost reduction.

\textbf{Freight by Road.} The primary metric for evaluating the improvement of predictive decision-making over reactive decision-making is the number of modal shifts from inland shipping to road transport. Such modal shifts lead to a disproportionate increase in both transportation costs and delays. Therefore, the predictive model minimizes the number of modal shifts, which consequently results in a higher total number of time steps due to conservative decision-making. This can, however, be mitigated by using a \textit{time to live} for individual ships, to ensure that ships also account for the fact that they should be transported quickly. 

\begin{table}[h]
\caption{Total distance covered over different models}
\label{tab:bottlenecks}
\centering

\setlength{\tabcolsep}{10pt}
\renewcommand{\arraystretch}{1.15}

\begin{tabular}{cccc}
\toprule
\textbf{Ships} & \textbf{Reactive} & \textbf{Predictive (70\%)} & \textbf{Predictive (90\%)} \\
\midrule
50  & 154 & 111 & 114 \\
100 & 287 & 206 & 203 \\
250 & 615 & 441 & 431 \\
500 & 851 & 821 & 811 \\
\bottomrule
\end{tabular}
\end{table}

\begin{figure}[h]
    \centering
    \includegraphics[width=1\linewidth]{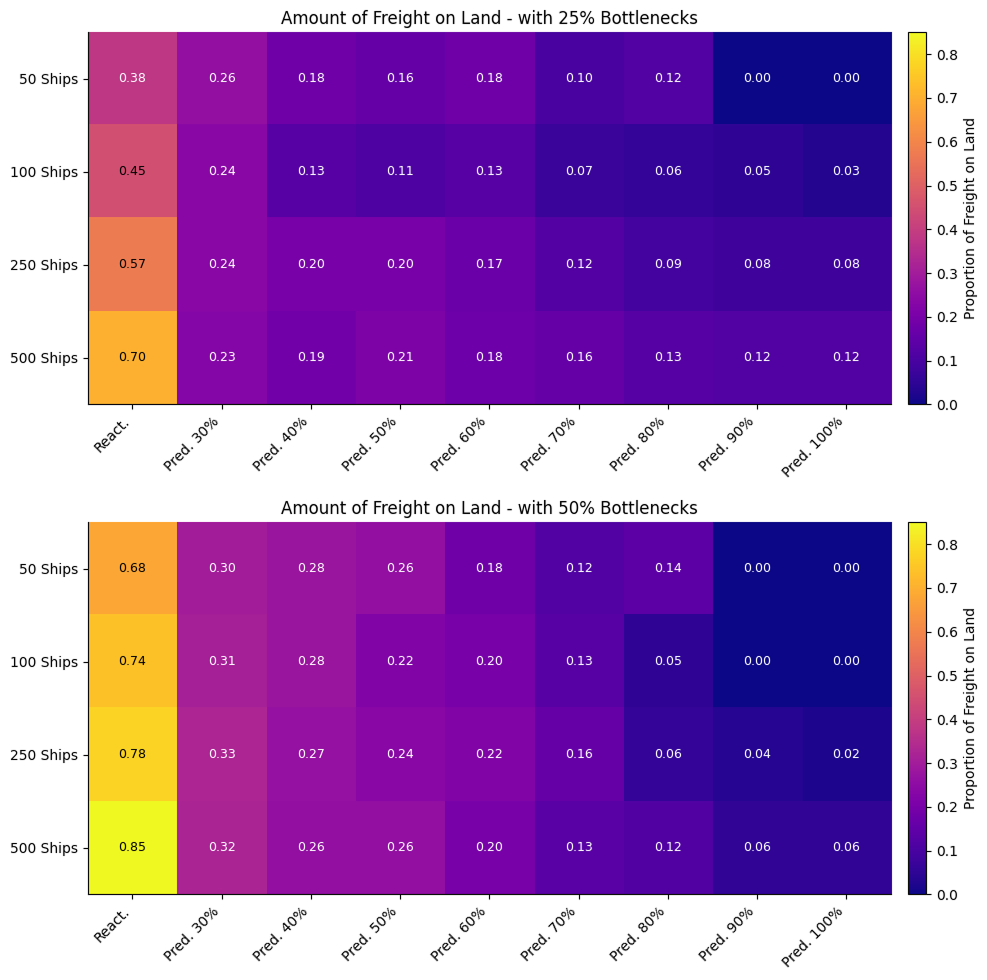}
    \caption{Freight on Land relative to the number of ships}
    \label{fig:25heatmap}
\end{figure}

In Figure \ref{fig:simulation}, we observe that a predictive model with an accuracy of 90\% reduces the number of modal shifts from 19 to zero. Furthermore, Figure \ref{fig:25heatmap} demonstrates that a predictive approach substantially decreases the relative number of modal shifts compared to using a reactive approach. This improvement remains visible even at moderate prediction accuracies. When comparing scenarios with increased bottlenecks, the predictive approach proves to be considerably more resilient than the reactive approach. However, as the number of bottlenecks increases, the importance of higher prediction accuracies becomes increasingly apparent. A similar trend appears when the number of ships increases.

It is important to note that by design, the predictive model can predict one arbitrary time step into the future, whereas the reactive model lacks this capability entirely. Consequently, even predictive models with relatively low accuracy tend to outperform the reactive approach, as they still retain a probability of correctly identifying low water levels in advance. As the prediction accuracy approaches 0\%, the model effectively behaves as a random classifier, assigning approximately equal probability to routes being classified as safe or unsafe. This explains the remaining discrepancy between low-accuracy predictive models and the reactive baseline. Nevertheless, within the accuracy range of 70\% to 100\%, the predictive model consistently demonstrates clear advantages, particularly under more extreme operating conditions.

\textbf{Steps taken by Ships.} Another relevant metric is the total number of steps taken by ships, or total distance covered, throughout the simulation. Table 2 shows that the reactive model consistently results in a substantially higher number of steps compared to the predictive model, even at a moderate prediction accuracy. Excluding the scenario with 500 ships, the predictive model achieves an average reduction of approximately 28.3\% in the total distance for accuracies between 70\% and 90\%. This reduction translates directly into lower fuel consumption and operational costs, as ships are required to travel shorter overall distances. However, as the operating conditions become more extreme, minimizing the amount of freight shifted to road transport requires increasingly conservative routing strategies. As a result, the reduction in total distance gradually diminishes as the waterway network becomes increasingly oversaturated.

When considering the reactive model, its performance is consistently lower due to the need for ad-hoc routing adjustments and modal shifts during operation to maintain efficient freight transportation. In contrast, the predictive model can account for these conditions at an earlier stage, allowing it to make more informed decisions proactively. As a result, the predictive approach reduces the operational overhead associated with last-minute rerouting and unnecessary modal shifts. 

\section{Expert Evaluation}
To validate the applicability of DTs in inland shipping, we conducted a series of semi-structured interviews with six experts \textit{(E1–E6)} with varied expertise in inland waterway logistics, resilience engineering, and policy making. Three participants were affiliated with academic institutions \textit{(E1– E3)}, two participants represented regional logistics and governmental organizations involved in inland freight and infrastructure coordination \textit{(E4, E5)}, and one participant was affiliated with industry-oriented resilience \textit{(E6)}. 

Each interview started with a discussion of the participant’s background and familiarity with DT systems, predictive decision-support, and resilience challenges. We subsequently introduced the DT concept investigated in this study and asked participants to provide their initial impressions regarding such a system. Depending on the participant’s expertise, interviews then focused on different aspects of the study. Domain experts in inland shipping and transportation evaluated the identified scenarios (i.e., Figure \ref{fig:scenarios}), response strategies, and the assumptions underlying the simulation model. Participants from related domains, such as resilience engineering, instead reflected on broader topics, including trust in predictive systems, stakeholder coordination, organizational adoption, and applicability in other domains. Finally, we discussed gaps in existing approaches, the potential impact of the DT, and recommendations for future development. The interview data were analyzed qualitatively to identify recurring themes, agreements, limitations, and possible extensions to both the DT and simulation model.

\textbf{General Digital Twin Concept.} The expert interviews generally supported the relevance of the proposed DT concept for inland waterway logistics, although several limitations and nuances were identified. Across both domain experts and experts from related fields, the DT was primarily viewed as a decision-support tool rather than an autonomous decision-maker. Several experts emphasized that its main value lies in consolidating fragmented information and combining real-time and predictive data to support operational decisions. At the same time, experts consistently stressed that human judgment should remain central, particularly because inland shipping decisions often depend on experience and contextual knowledge. Concerns regarding over-reliance on automated recommendations were repeatedly raised, leading several experts to position the DT primarily as a tool for simulations and scenario exploration rather than full operational automation.

\textbf{Scenario Analysis.} Multiple domain experts agreed that over-saturation, alarming water levels, and blocked waterways capture the primary disruption categories within inland waterway logistics. However, experts also highlighted additional bottlenecks unrelated to water levels, particularly infrastructure failures such as malfunctioning locks and bridges. Regarding responses, rerouting and modal shifts toward road transport were described as the most common approaches currently used in practice, while cargo splitting and smaller vessels are used to a lesser extent due to infrastructural and organizational limitations. Experts additionally emphasized that companies already operate semi-proactively by relying on manual intervention using weather forecasts and seasonal patterns, although stakeholders still expressed a need for more integrated and reliable predictive support.

\textbf{Application of DT Systems in Freight.} A recurring theme throughout the interviews concerned the practical role of DT systems within inland freight logistics. Most experts agreed that predictive decision-support systems provide great value, but several noted that inland shipping is non-volatile and may not require continuous real-time decision-making, as DTs are often used for. One expert specifically described inland freight as slow-moving and schedule-driven compared to highly volatile domains. Consequently, experts emphasized that the greatest value of the DT lies in providing a digital shadow to support operational planning, trade-off evaluation, and scenario analysis through simulations rather than continuously automating decisions. Moreover, several experts stressed that successful adoption depends not only on technical sophistication but also on interpretability and stakeholder trust.

\textbf{Simulation Model and Parameters.} The simulation model and its parameters were considered an appropriate abstraction for evaluating DT-supported logistics decisions. The graph-based structure was viewed as representative for simulation purposes, although multiple experts suggested possible extensions. In particular, experts highlighted the importance of validating the model using historical operational data and incorporating additional variables such as weather conditions, shipment urgency, and emissions trade-offs. One expert emphasized that the practical usefulness of the simulation depends on its ability to reproduce historical patterns. Overall, the interviews strongly supported the model while also identifying several opportunities for future refinement.

\section{Discussion}
\textbf{Opportunities.} The results suggest that even moderate predictive capabilities, enabled by DTs, can substantially improve resilience in inland freight logistics compared to purely reactive approaches. Across nearly all evaluated settings, the predictive approach reduced modal shifts toward road transport while reducing the total distance covered by ships. This is important, as modal shifts toward road transport generally represent the most expensive and environmentally undesirable outcome. Interestingly, although the predictive model often required more time steps overall due to proactive rerouting and precautionary interventions, ships frequently followed more efficient routes, reducing unnecessary movements and therefore fuel costs. This suggests that DT-supported predictive logistics in inland shipping may not only reduce disruptions but also improve operational efficiency and sustainability.

Moreover, the simulation model and expert study show the applicability of a DT in the context of inland waterway logistics. Despite the non-volatile nature, the DT enables predictive simulations and, by extension, decision-making by storing real-time data in a digital shadow. The expert interviews reinforced that inland freight logistics is a non-volatile environment, compared to one that requires constant real-time optimization. As a result, the primary value of the DT may lie less in autonomous operational control and more in supporting anticipatory planning and decision-making, as opposed to volatile settings. In this context, the DT functions as a decision-support environment based on a digital shadow of the waterway system. Experts repeatedly emphasized that the ability to evaluate trade-offs and explore possible future outcomes may be more valuable than automation itself.

Another key opportunity lies in stakeholder coordination. The simulation shows that predictive decision-making reduces modal shifts and improves routing efficiency under congestion, suggesting clear benefits of shared anticipatory information. Despite extreme conditions, the predictive approach showed clear improvements over a reactive approach. Combined with expert feedback on fragmented data across stakeholders, this indicates that Digital Twins can enhance transparency and coordination in inland freight logistics, even while preserving human-centered decision-making, as decisions made by stakeholders will be reflected in the DT.

\textbf{Limitations and Future Work.} Although the simulation results and expert evaluations support the general direction of the proposed DT approach, several limitations and extensions were identified. Most importantly, multiple experts stressed the importance of validating the simulation against historical operational and environmental data. While the current abstraction allows for generalization and is grounded in real-world logistics behavior and expert discussions, future work should investigate whether similar trends persist when using actual inland waterway data, weather patterns, infrastructure conditions, and historical disruptions.

Several additional extensions were also identified during the expert interviews. These include incorporating variables such as shipment urgency, inventory constraints, infrastructure failures, emissions trade-offs, and more advanced optimization mechanisms. Moreover, experts noted that the abstract nature of the simulation model makes it generalizable and potentially applicable beyond inland waterways. Similar approaches may therefore be explored in domains such as road-based freight logistics, supply chain resilience, infrastructure monitoring, energy systems, or disaster management. Finally, future DT implementations may require stakeholder-specific interfaces and recommendation systems tailored toward different operational goals, incentives, and levels of technical expertise, as a primary challenge in inland waterway logistics is integrating data from different stakeholders.

\section{Conclusion}
While Digital Twins (DT) are traditionally used for volatile, fast-moving scenarios, we have shown through simulations that they can be useful for decision-making in non-volatile contexts, such as inland waterway logistics. Our work shows that even with moderate accuracy, a predictive approach can provide a substantial improvement over a reactive approach, despite facing extreme conditions such as a large number of ships or increased bottlenecks.

According to the interviewed domain experts, using DTs can improve decision-making in inland freight shipping, as there exist many stakeholders with differing preferences and interests. By combining predictive insights, simulation-based scenario exploration, and shared operational information, a DT can provide a common and powerful decision-support environment that improves both coordination and resilience while operating on real-time data. 

Overall, our results indicate that DT-supported predictive logistics can reduce modal shifts, lower operational overhead, and improve the robustness of inland freight transportation under increasing climate-related uncertainty. Although further validation using historical data remains a next step, this work demonstrates the potential of DTs as a practical decision-support tool for resilient inland waterway logistics.

\section*{Acknowledgments}

The authors would like to thank Anne-Ruth Scheijgrond, Martijn Mes, Davoud Hosseinezhad, Stephanie Hessing, Christa Baas, Joschka Hullmann and Tobias Stähle for providing valuable feedback to this work. Additionally, the authors thank the six domain experts for participating in the interviews.

\bibliographystyle{IEEEtran}
\bibliography{references}

\end{document}